\documentclass[fleqn,aps,pra,amsmath,amssymb,preprint]{revtex4-1}

\usepackage{float}
\usepackage{graphicx}
\usepackage{textcomp}
\usepackage{braket}
\usepackage{enumerate}
\usepackage{multirow}
\usepackage{array}
\usepackage{tabularx}

\renewcommand\Im{\operatorname{Im}}

\begin{document}

\title{Role of dressed-state interference in electromagnetically induced transparency}
 
 \author{Sumanta Khan}
 \affiliation{Department of Physics, Indian Institute of
 Science, Bangalore 560\,012, India}

 \author{Vineet Bharti}
 \affiliation{Department of Physics, Indian Institute of
 Science, Bangalore 560\,012, India}

\author{Vasant Natarajan}
\affiliation{Department of Physics, Indian Institute of
 Science, Bangalore 560\,012, India}
 \email{vasant@physics.iisc.ernet.in}
 \homepage{www.physics.iisc.ernet.in/~vasant}

\begin{abstract}
Electromagnetically induced transparency (EIT) in three-level systems uses a strong control laser on one transition to modify the absorption of a weak probe laser on a second transition. The control laser creates dressed states whose decay pathways show interference. We study the role of dressed-state interference in causing EIT in the three types of three-level systems -- lambda ($\Lambda$), ladder ($\Xi$), and vee (V). In order to get realistic values for the linewidths of the energy levels involved, we consider appropriate hyperfine levels of $^{87}$Rb. For such realistic systems, we find that dressed-state interference causes probe absorption---given by the imaginary part of the susceptibility---to go to zero in a $\Lambda$ system, but plays a negligible role in $\Xi$ and V systems.

\noindent
\textbf{Keywords}: Electromagnetically induced transparency; Coherent control; Quantum optics.

\end{abstract}

\maketitle

\section{Introduction}
Electromagnetically induced transparency (EIT) \cite{BIH91,FIM05} is a phenomenon in which a strong control laser is used to modify the properties of a medium for a weak probe laser. The phenomenon uses the fact that most optical media are composed of multilevel atoms, so that the control and probe beams can act on different transitions. Applications of EIT include slowing of light \cite{HHD99} (for use in quantum-information processing), lasing without inversion \cite{AGA91}, enhanced nonlinear optics \cite{HFI90,ZAX08}, high-resolution spectroscopy \cite{RAN02,DAN05,KPW05}, and getting subnatural linewidth for tight locking of lasers to optical transitions \cite{RWN03,IKN08}.

EIT occurs due to two effects caused by the strong control laser -- (i) AC Stark shift of the atomic levels leading to the creation of new eigenstates of the coupled atom plus photon system called {\it dressed states} (see Refs.\ \cite{COR77a,COR77}; but with the interaction Hamiltonian included as described in Ref.\ \cite{FIM05}), and (ii) the interference between the decay pathways to or from these dressed states. The dressed-state approach is better because it is valid at all intensities (page 98 of Ref.\ \cite{NAT15}), whereas the one used in Ref.\ \cite{GOA80} is valid only at high intensities. The degree of interference depends on the linewidth of the dressed states, and the role of this interference in EIT is not well studied in all three-level systems. 

In this work, we study in detail the dressed-state linewidth and interference in the three canonical types of three-level atoms---lambda ($\Lambda$), ladder ($\Xi$), and vee (V). For specificity, we choose hyperfine energy levels of $^{87}${Rb}. This allows us to use realistic values of the decay rates, which is important since it influences the degree of interference in $\Xi$ and V systems that involve multiple excited states. We find that the dressed-state linewidth steadily increases over these three systems. In addition, interference causes probe absorption to vanish identically at line center in $\Lambda$ systems because of the formation of a dark state. As expected, \textit{the effect of interference decreases with increasing Rabi frequency of the control laser, since it causes increasing separation of the dressed states}.

\section{Dressed-state location and linewidth}
We define the three atomic levels as $\ket{1}$, $\ket{2}$, and $\ket{3}$. The probe laser drives the $\ket{1} \leftrightarrow \ket{2}$ transition with Rabi frequency $\Omega_p$ and detuning $\Delta_p$. The strong control laser drives the $\ket{2} \leftrightarrow \ket{3}$ transition with Rabi frequency $\Omega_c$ and detuning $\Delta_c$. Both these transitions are electric dipole (E1) allowed, and the levels involved have opposite parity. Therefore, the $\ket{1} \leftrightarrow \ket{3}$ transition involves levels of the same parity and is E1 forbidden. The level $\ket{i}$ is assumed to have a spontaneous decay rate of $\Gamma_i$. Since ground levels have zero decay rate, depending on the particular three-level system, one or more of the $\Gamma_i$'s will be zero. 

\begin{figure}
	\centering{\resizebox{0.99\columnwidth}{!}{\includegraphics{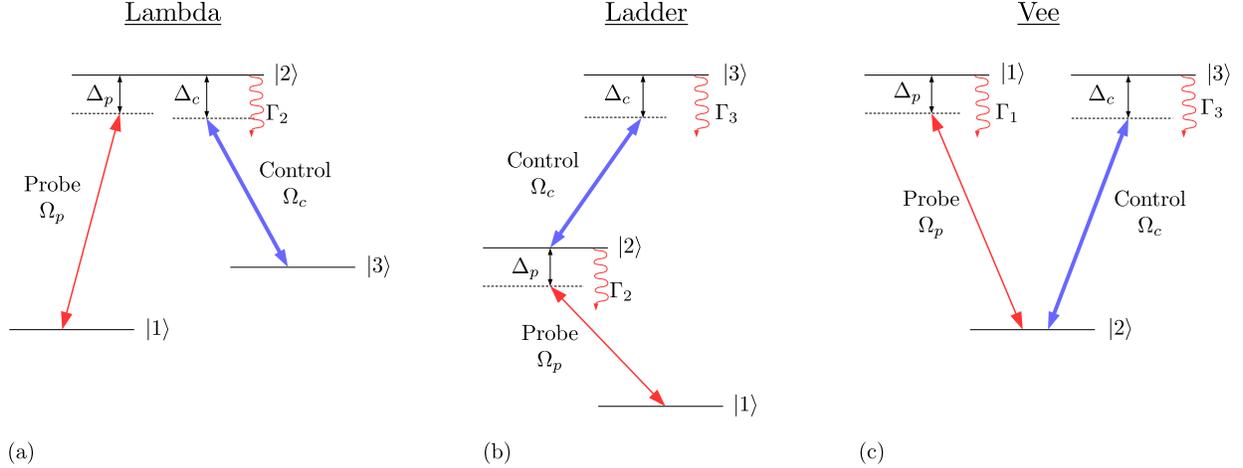}}}
	\caption{(Color online) Three-level systems. (a) $\Lambda$ system. (b) $\Xi$ system. (c) V system.}
	\label{model}
\end{figure}

The 3 three-level systems are shown in Fig.\ \ref{model}. Theoretical analysis for each system is done using a standard density matrix analysis of the three levels involved. Time evolution of the elements is given by the general equation:
\begin{equation}
\dfrac{d\rho}{dt} = -\,\dfrac{i}{\hbar} \left[H,\rho\right] - \dfrac{1}{2}\{\Gamma,\rho\}
\end{equation}
where $H$ is the Hamiltonian governing the atom-photon interaction, and $\Gamma$ is the relaxation rate. The coherence induced on the probe transition is given by the element $\rho_{12}$. Therefore, probe absorption is proportional to its imaginary part, while probe dispersion is proportional to its real part. In the following, we will operate in the weak-probe limit, i.e.\ $\Omega_p \ll \Omega_c$. 


The dressed states created by the control laser are shifted in energy by the AC stark shift. Therefore they are located at values of $ \Delta_p $ given by:
\begin{equation}
\Delta_{\pm} = \dfrac{\Delta_c}{2} \pm \dfrac{1}{2} \sqrt{\Delta^2_c + \Omega^2_c} 
\end{equation}
If we consider the special case of $ \Delta_c=0 $, i.e.\ control laser on resonance, then probe absorption splits into the expected Autler-Townes doublet with peaks located at $ \pm \Omega_c/2 $. Since the control laser couples levels $ \ket{2} $ and $ \ket{3} $, the linewidth of each dressed state is given by:
\begin{equation}
\Gamma_{\rm ds} = \dfrac{\Gamma_2 + \Gamma_3}{2}
\end{equation}
The probe laser couples levels $ \ket{1} $ and $ \ket{2} $, hence the linewidth of each sub-peak in the probe absorption spectrum is:
\begin{equation}
\Gamma_p = \Gamma_1 + \dfrac{\Gamma_2 + \Gamma_3}{2}
 \label{eq:dressedwidth}
\end{equation}
In the following, the above expression will be used to determine the linewidth of the Autler-Townes doublet in the 3 three-level systems.

\section{EIT in a lambda system}
We first consider the $\Lambda$ system shown in Fig.\ \ref{model}(a). The energy levels of $^{87}$Rb used to form this system are shown in the table below.
\begin{displaymath}
\begin{array}{cc}
\hline \hline \\
\begin{array}{>{\centering\arraybackslash}p{4em}>{\centering}p{4cm}>{\centering\arraybackslash}p{1.5cm}}
Level\hspace*{4mm} & $^{87}$Rb hyperfine level\hspace*{4mm} & $\Gamma/2\pi\hspace*{4mm}$ \\ 
& &\hspace*{-4mm}(MHz)\\ 
\end{array}
&
\begin{array}{>{\centering}p{2.7cm}}
Wavelength\\
(nm)
\end{array}
\\
\hline  \\[-1.2em]
\left.
\begin{array}{>{\centering}p{4em}>{\centering}p{4cm}>{\centering\arraybackslash}p{1.5cm}}
\\ [-0.7 em]
$ \ket{1} $ & $ \rm 5S_{1/2},F=1 $ & $ 0 $ \vspace*{0.5em}\\
$ \ket{2} $ & $ \rm 5P_{3/2}, F=2 $ & $ 6.1 $ \vspace*{0.5em}\\
$ \ket{3} $ & $ \rm 5S_{1/2}, F=2 $ & $ 0 $ \vspace*{0.5em}\\ [-0.1em]
\end{array}
\right\} 
& 
\begin{array}{>{\centering\arraybackslash}p{2.7cm}}
\multirow{3}{*}{\parbox[t]{3cm}{\centering $ \lambda_p=780.24$ $\lambda_c=780.24 $}}\\
\\
\\
\end{array}\\
\hline
\end{array}
\end{displaymath}
As seen, $\ket{2}$ is an excited state, and $\ket{1}$ and $\ket{3}$ are ground levels. Therefore, $\Gamma_1 = \Gamma_3 = 0$ and $\Gamma_2$ is non-zero. 

\begin{figure}
	\centering{\resizebox{0.6\columnwidth}{!}{\includegraphics{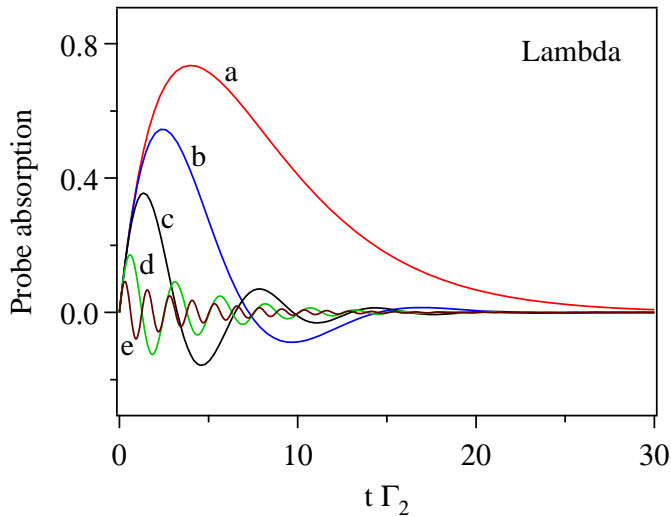}}}
	\caption{(Color online) Transient behavior of probe absorption, given by $ \Im \{ \rho_{12}\Gamma_2/\Omega_p\} $, for a $ \Lambda $ system. Spectra are shown for $ \Delta_p = \Delta_c = 0 $, and five values of $\Omega_c$: $\rm a=0.5 \, \Gamma_2$, $\rm b= 1 \, \Gamma_2$, $\rm c= 2 \, \Gamma_2$, $\rm d = 5 \, \Gamma_2$, $\rm e = 10 \, \Gamma_2 $.}
	\label{lambda_transient}
\end{figure}

In order to ensure a steady state approach ($\dot\rho=0) $ is valid, we have studied the transient behavior of the density-matrix elements of this system. Probe absorption, which is given by $ \Im \{ \rho_{12}\Gamma_2/\Omega_p\} $, is plotted in Fig.\ \ref{lambda_transient} for 5 values of $ \Omega_c $ (negative absorption means gain). It is clear that transient oscillations die down after a few lifetimes $ \Gamma_2 $ of the excited state. Thus, the system will reach steady state within the time frame used in the experiment, and the density-matrix equations can be solved under this condition. Such an analysis yields \cite{LIX95R}:
\begin{equation}
\begin{aligned}
\rho_{12} &= \dfrac{i \Omega_{p}/2}{\left(\dfrac{\Gamma_2}{2}-i\Delta_{p}\right)+ \displaystyle{\dfrac{i|\Omega_{c}|^{2}/4}{\left(\Delta_{p}-\Delta_{c}\right)}}}
\label{lambdasol}
\end{aligned}
\end{equation}
From Eq.\ \eqref{eq:dressedwidth}, we get the linewidth of the Autler-Townes doublet as 
\begin{equation}
\begin{aligned}
\Gamma_{\rm AT} = \dfrac{\Gamma_2}{2}
\end{aligned}
\end{equation}
Thus, probe absorption splits into two peaks, each with a linewidth of half the original linewidth.

\begin{figure}
\centering{\resizebox{0.4\columnwidth}{!}{\includegraphics{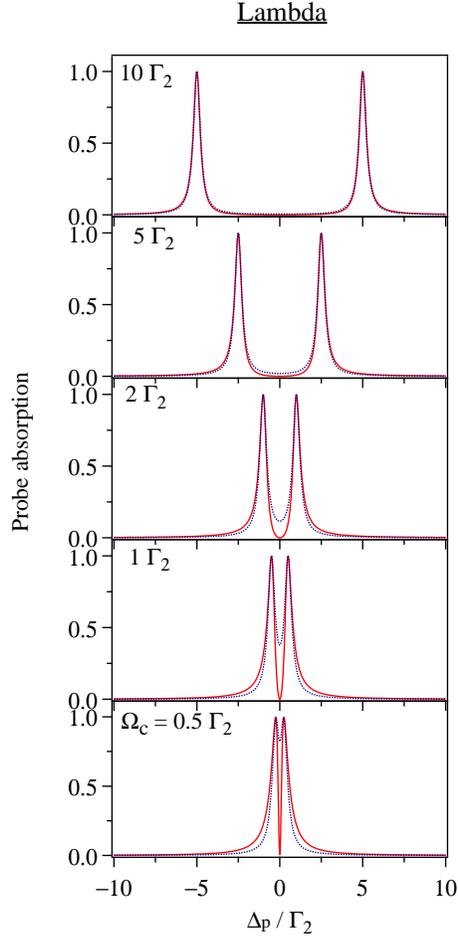}}}
\caption{(Color online) EIT in the $\Lambda$ system. Probe spectra are shown for $\Delta_c=0$ and five values of $\Omega_c$. The solid lines represent the complete density-matrix calculation, while the dashed lines represent what the spectrum would be with two dressed states of linewidth $\Gamma_2/2$ and separation $\Omega_c$ (no interference).}
 \label{lambdaeitvslorz}
\end{figure}

The above description ignores any interference between the decay pathways to the two dressed states. If we take that into account, there is destructive interference at line center and probe absorption goes identically to zero. For low values of control Rabi frequency $\Omega_c \ll \Gamma_2$, this interference can make the EIT dip extremely narrow. Such dressed-state interference in a $\Lambda$ system has been observed before using the $\rm D_1$ line of Rb \cite{LIX95}. The effect of this interference is seen clearly in our calculations shown in Fig.\ \ref{lambdaeitvslorz}. The solid lines represent the complete density-matrix calculation, while the dashed lines are what the spectrum would look like with just the two dressed states of linewidth $\Gamma_2/2$ and no interference. A quantitative measure of what is shown in the figure is seen in the table below (at line center, $ \Delta_p = 0 $). The last column, which gives the difference between having and not having interference, shows that the effect of interference decreases as the control Rabi frequency is increased.

\begin{center}
\begin{tabular}{c c c c}
\hline \hline
$\Omega_c$ & No interference  & With interference & Difference   \\
$(\Gamma_2) $ & (\%)  & (\%) & (\%)\\  
\hline \\[-0.5em]
0.5 & 0.8235 &  0.0064 & 0.8171  \\
1 & 0.3778 &  0.0004 & 0.3774 \\
2 & 0.1159 &  0 & 0.1159  \\
5 & 0.0197 &  0 & 0.0197 \\
10 & 0.0049 &  0 & 0.0049  \\
\hline \hline  
\end{tabular}
\end{center}

Thus, EIT in a $\Lambda$ system is explained by a combination of the AC Stark shift caused by the control laser and the interference of the decay pathways from the dressed states, especially for small values of control Rabi frequency.

\section{EIT in a ladder system}
We now consider the $\Xi$ system shown in Fig.\ \ref{model}(b). The energy levels of $^{87}$Rb used to form this system are shown in the table below.
\begin{displaymath}
\begin{array}{cc}
\hline \hline \\
\begin{array}{>{\centering\arraybackslash}p{4em}>{\centering}p{4cm}>{\centering\arraybackslash}p{1.5cm}}
Level\hspace*{4mm} & $^{87}$Rb hyperfine level\hspace*{4mm} & $\Gamma/2\pi\hspace*{4mm}$ \\ 
& &\hspace*{-4mm}(MHz)\\ 
\end{array}
&
\begin{array}{>{\centering}p{2.7cm}}
Wavelength\\
(nm)
\end{array}
\\
\hline  \\[-1.2em]
\left.
\begin{array}{>{\centering}p{4em}>{\centering}p{4cm}>{\centering\arraybackslash}p{1.5cm}}
\\ [-0.7 em]
$ \ket {1} $ & $ \rm 5S_{1/2},F=1 $ & $ 0 $ \vspace*{0.5em}  \\
$ \ket{2} $ & $ \rm 5P_{3/2}, F=2 $ & $ 6.1 $ \vspace*{0.5em}  \\
$ \ket{3} $ & $ \rm 5D_{5/2}, F=1 $ & $ 0.68 $ \vspace*{0.5em}  \\ [-0.1em]
\end{array}
\right\} 
& 
\begin{array}{>{\centering\arraybackslash}p{2.7cm}}
\multirow{3}{*}{\parbox[t]{3cm}{\centering $ \lambda_p=780.24$ $\lambda_c=775.80 $}}\\
\\
\\
\end{array}\\
\hline
\end{array}
\end{displaymath}
Here, the two levels coupled by the control are both excited, therefore $\Gamma_2$ and $\Gamma_3$ are non-zero and $\Gamma_1 = 0$. 

\begin{figure}
	\centering{\resizebox{0.6\columnwidth}{!}{\includegraphics{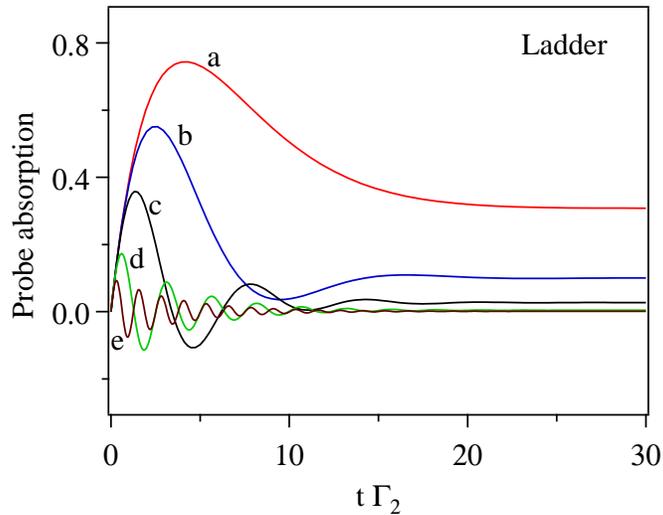}}}
	\caption{(Color online) Transient behavior of probe absorption, given by $ \Im \{ \rho_{12}\Gamma_2/\Omega_p\} $, for a $ \Xi $ system. Spectra are shown for $ \Delta_p = \Delta_c = 0 $, and five values of $\Omega_c$: $\rm a=0.5 \, \Gamma_2$, $\rm b= 1 \, \Gamma_2$, $\rm c= 2 \, \Gamma_2$, $\rm d = 5 \, \Gamma_2$, $\rm e = 10 \, \Gamma_2 $.}
	\label{ladder_transient}
\end{figure}

As before, we study the transient behavior to verify that a steady state solution is valid. Probe absorption spectra, given by $ \Im \{ \rho_{12} \Gamma_2/\Omega_p \} $ and plotted in Fig.\ \ref{ladder_transient}, show that all transient oscillations die down after a few lifetimes $ \Gamma_2 $ of the excited state. Thus the density-matrix analysis can be done in steady state, which yields a solution of \cite{KPW05}:
\begin{equation}
\begin{aligned}
\rho_{12} &= \dfrac{i \Omega_{p}/2}{\left(\dfrac{\Gamma_{2}}{2}-i\Delta_{p}\right)+ \displaystyle{\dfrac{|\Omega_{c}|^{2}/4}{\left(\dfrac{\Gamma_3}{2}-i(\Delta_{p}+\Delta_{c})\right)}}}
\label{laddersol}
\end{aligned}
\end{equation}

\begin{figure}
	\centering{\resizebox{0.4\columnwidth}{!}{\includegraphics{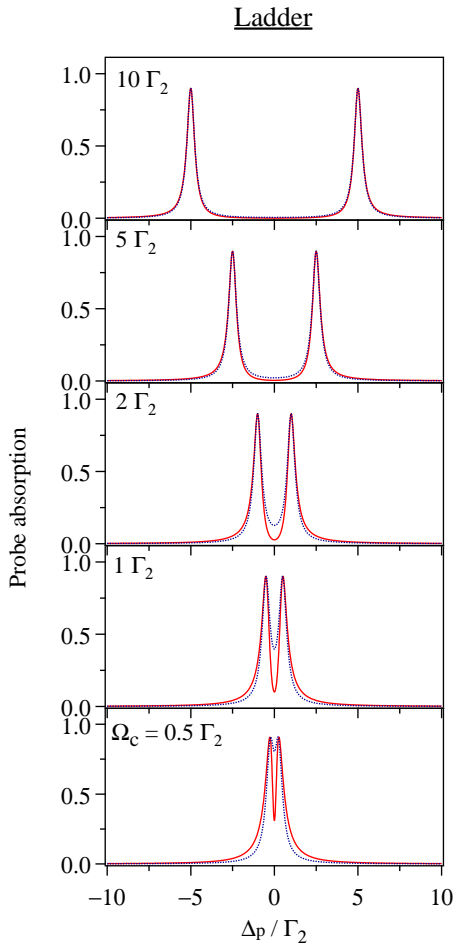}}}
	\caption{(Color online) EIT in the $\Xi$ system. Probe spectra are shown for $\Delta_c=0$ and five values of $\Omega_c$. The solid lines represent the complete density-matrix calculation, while the dashed lines represent what the spectrum would be with two dressed states of linewidth $(\Gamma_2+\Gamma_3)/2$ and separation $\Omega_c$ (no interference).}
	\label{ladder_actualdecayrate}
\end{figure}

EIT spectra are shown in Fig.\ \ref{ladder_actualdecayrate}. Again taking the special case of control on resonance $\Delta_c=0$, we see from the figure that each  spectrum splits into two (Autler-Townes doublet) with a transparency dip at line center. From Eq.\ \eqref{eq:dressedwidth}, the linewidth of each sub-peak is:
\begin{equation}
\begin{aligned}
\Gamma_{\rm AT} = \dfrac{\Gamma_2+\Gamma_3}{2}
\end{aligned}
\end{equation}

But in this case the absorption at line center does not go to zero, because no stable dark state is formed. In addition, interference causes the transparency dip to increase from that of the Autler-Townes doublet, but this is peculiar to the linewidth of the upper state in $^{87}\rm Rb $. In fact, it has been shown that the degree of interference becomes zero (no difference between solid and dashed lines in the figure) when $ \Gamma_3$ becomes equal to $ \Gamma_2 $ \cite{AGA97}. A quantitative picture of what is shown in the figure is seen in the table below (at line center, $ \Delta_p = 0 $). The degree of interference is smaller than in the $ \Lambda $ system, but as before the effect of interference decreases with increasing $ \Omega_c $
\begin{center}
	\begin{tabular}{c c c c}
		\hline \hline
		$\Omega_c$  & No interference  & With interference  & Difference  \\ 
		$(\Gamma_2) $ & (\%)  & (\%) & (\%)\\  
		\hline \\[-0.5em]
		0.5 & 0.8062 &  0.3109 & 0.4953  \\
		1 & 0.3975 &  0.1006 & 0.2969 \\
		2 & 0.1268 &  0.0271 & 0.0997  \\
		5 & 0.0219 &  0.0044 & 0.0175 \\
		10 & 0.0055 &  0.0011 & 0.0044  \\
		\hline \hline  
	\end{tabular}
\end{center}

Thus, we conclude that EIT in a ladder system can be explained mainly by the AC stark shift caused by the control laser, and depends only slightly on dressed-state interference.

\section{EIT in a vee system}
We finally consider the V system shown in Fig.\ \ref{model}(c). The energy levels of $^{87}$Rb used to form this system are shown in the table below.
\begin{displaymath}
\begin{array}{cc}
\hline \hline \\
\begin{array}{>{\centering\arraybackslash}p{4em}>{\centering}p{4cm}>{\centering\arraybackslash}p{1.5cm}}
Level\hspace*{4mm} & $^{87}$Rb hyperfine level\hspace*{4mm} & $\Gamma/2\pi\hspace*{4mm}$ \\ 
& &\hspace*{-4mm}(MHz)\\ 
\end{array}
&
\begin{array}{>{\centering}p{2.7cm}}
Wavelength\\
(nm)
\end{array}
\\
\hline  \\[-1.2em]
\left.
\begin{array}{>{\centering}p{4em}>{\centering}p{4cm}>{\centering\arraybackslash}p{1.5cm}}
\\ [-0.7 em]
$ \ket {1} $ & $ \rm 5P_{3/2},F=1 $ & $ 6.1 $ \vspace*{0.5em}  \\
$ \ket{2} $ & $ \rm 5S_{1/2}, F=1 $ & $ 0 $ \vspace*{0.5em}  \\
$ \ket{3} $ & $ \rm 5P_{3/2}, F=2 $ & $ 6.1 $ \vspace*{0.5em}  \\ [-0.1em]
\end{array}
\right\} 
& 
\begin{array}{>{\centering\arraybackslash}p{2.7cm}}
\multirow{3}{*}{\parbox[t]{3cm}{\centering $ \lambda_p=780.24$ $\lambda_c=780.24 $}}\\
\\
\\
\end{array}\\
\hline
\end{array}
\end{displaymath}
Here, the common level is a ground level, and both the control and probe lasers couple to excited levels. Therefore, $\Gamma_1$ and $\Gamma_3$ are non-zero and $\Gamma_2 = 0$. In this case, probe absorption is given by $ \Im \{ \rho_{21} \Gamma_1/\Omega_p \} $. 

\begin{figure}
	\centering{\resizebox{0.6\columnwidth}{!}{\includegraphics{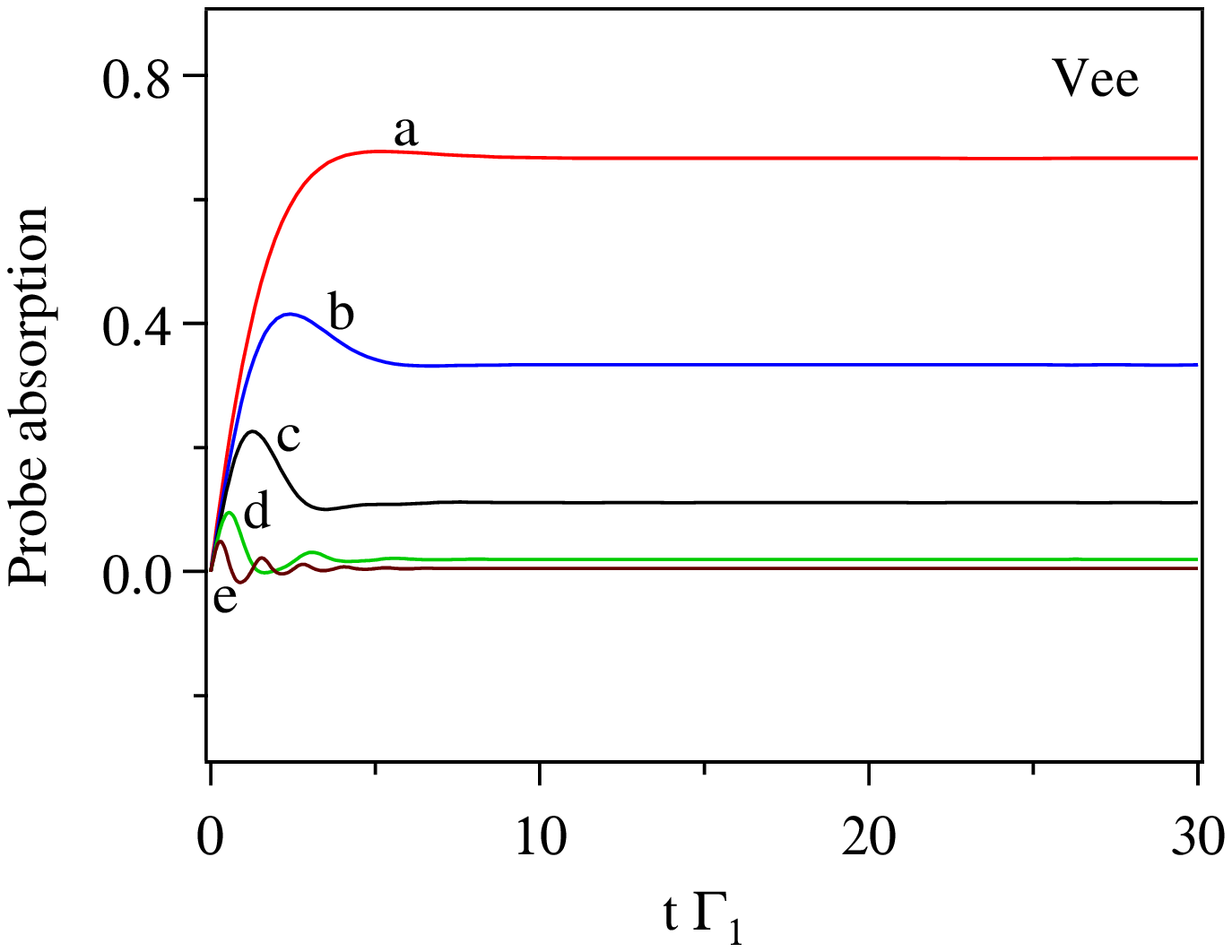}}}
	\caption{(Color online) Transient behavior of probe absorption, given by $ \Im \{ \rho_{21}\Gamma_1/\Omega_p\} $, for a V system. Spectra are shown for $ \Delta_p = \Delta_c = 0 $, and five values of $\Omega_c$: $\rm a=0.5 \, \Gamma_1$, $\rm b= 1 \, \Gamma_1$, $\rm c= 2 \, \Gamma_1$, $\rm d = 5 \, \Gamma_1$, $\rm e = 10 \, \Gamma_1 $.}
	\label{vee_transient}
\end{figure}

Transient spectra, shown in Fig.\ \ref{vee_transient}, show that the system reaches steady state after a few lifetimes $ \Gamma_1 $ of the excited state. Hence this system is also analyzed in steady state, which yields a solution of \cite{RAN04}:
\begin{equation}
\begin{aligned}
\rho_{21} &= \dfrac{i\Omega_{p}\left[ (\Gamma_3^2+4\Delta^{2}_{c}+|\Omega_{c}|^{2})-\dfrac{|\Omega_{c}|^{2}\left(\dfrac{\Gamma_3}{2}-i\Delta_{c}\right)}{\left(\dfrac{\Gamma_1+\Gamma_3}{2}
		+i(\Delta_{c}-\Delta_{p})\right)}\right]}{2(\Gamma_3^2+4\Delta_c^2+2|\Omega_{c}|^{2})\left[\dfrac{\Gamma_1}{2}-i\Delta_{p}
	+\dfrac{|\Omega_{c}|^{2}/4}{\left(\dfrac{\Gamma_1+\Gamma_3}{2}+i(\Delta_{c}-\Delta_{p})\right)}\right]}
\label{veesol}
\end{aligned}
\end{equation}

\begin{figure}
	\centering{\resizebox{0.4\columnwidth}{!}{\includegraphics{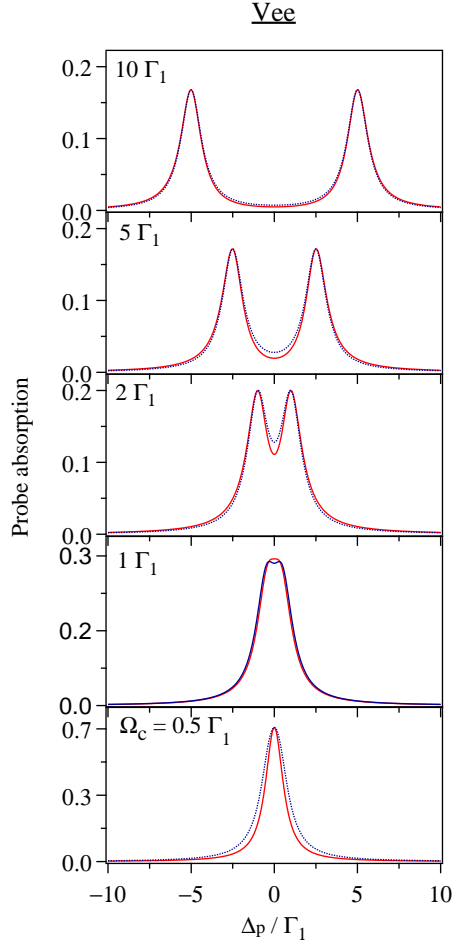}}}
	\caption{(Color online) EIT in the V system. Probe spectra are shown for $\Delta_c=0$ and five values of $\Omega_c$. The solid lines represent the complete density-matrix calculation, while the dashed lines represent what the spectrum would be with two dressed states of linewidth $( \Gamma_1 + \Gamma_3/2 )$ and separation $\Omega_c$ (no interference).}
	\label{veeeitvslorz}
\end{figure}

EIT spectra are shown in Fig.\ \ref{veeeitvslorz} for the case of $\Delta_c=0$ and 5 values of $\Omega_c$. As in the other cases, the spectrum splits into an Autler-Townes doublet. The linewidth of each doublet peak from Eq.\ \eqref{eq:dressedwidth} is:
\begin{equation}
\begin{aligned}
\Gamma_{\rm AT} = \Gamma_1 + \dfrac{\Gamma_3}{2}
\end{aligned}
\end{equation}

\noindent
The spectra calculated with and without interference show that, as in the case of the ladder system, the effect of interference is quite small. In addition, EIT is only partial, because no dark state can be formed. The table below shows this behavior quantitatively (at line center, $ \Delta_p = 0 $). The first two rows have negative difference values, which means that the interference is constructive and something that is different from all other cases, both in this system and other systems. 
\begin{center}
	\begin{tabular}{c c c c}
		\hline \hline
		$\Omega_c$ & No interference & With interference  & Difference   \\ 
		$(\Gamma_1) $ & (\%)  & (\%) & (\%)\\  
		\hline \\[-0.5em]
		0.5 & 0.6664 &  0.6665 & $-$0.0001  \\
		1 & 0.3235 &  0.3333 & $-$0.0098 \\
		2 & 0.1281 &  0.1111 & 0.0170  \\
		5 & 0.0277 &  0.0196 & 0.0081 \\
		10 & 0.0073 &  0.0049 & 0.0024  \\
		\hline \hline  
	\end{tabular}
\end{center}

Thus, EIT in a V system is almost exclusively dependent on the AC Stark shift with negligible role for dressed-state interference.

\section{Conclusion}
In summary, we have studied the linewidth of the dressed states and the role of interference in the three types of three-level systems. All the three show an Autler-Townes doublet structure in the probe-absorption spectrum. This results in a transparency window at line center, which is the main cause for EIT in these systems. However, dressed-state interference plays an equally important role in the phenomenon of EIT in a lambda system. This causes the EIT dip to go identically to zero at line center---this can result in an extremely narrow transparency window when the control power is very small. By contrast, dressed-state interference plays a negligible role in ladder and vee systems.

\section*{Acknowledgments}
This work was supported by the Department of Science and Technology, India. S K acknowledges financial support from an INSPIRE Fellowship, Department of Science and Technology, India; and V B from a D S Kothari post-doctoral fellowship of the University Grants Commission, India. The authors thank S Raghuveer for help with the manuscript preparation.


%

\end{document}